\documentstyle[draft]{mn}

\catcode`\"=\active\let"=\"

\def\3{{\ss} }

\def\c12{{1\over 2}}

\def\G{{\rm G}}
\def\d{{\rm d}}

\def\plusplus{\raise 0.3ex\hbox{${\scriptstyle ++}$}{}}

\newcommand{\oversim}[2]{\protect{\mbox{\lower0.5ex\vbox{%
   \baselineskip=0pt\lineskip=0.2ex
   \ialign{$\mathsurround=0pt #1\hfil##\hfil$\crcr#2\crcr\sim\crcr}}}}} 
\newcommand{\simless} {\mbox{$\,\mathrel{\mathpalette\oversim<}\,$}} % <~ sign

\title[Effects of dynamical evolution on the distribution of substructures] {Effects of dynamical evolution on the distribution of substructures}
\author[J. Pe\~{n}arrubia \& A. J. Benson ]
{Jorge Pe\~{n}arrubia$^1$
, Andrew J. Benson$^2$\\
$^1$Max Planck Institute f{\"u}r Astronomie, K{\"o}nigstuhl 17, D-69117, Germany \\
$^2$Department of Physics, University of Oxford, Keble Road, Oxford, OX1 3RH, U.K.\\}

\begin{document}

\maketitle

\begin{abstract}
We develop a semi-analytical model that determines the evolution of the mass, position and internal structure
of dark matter substructures orbiting in dark matter haloes. We apply this model to the case of the Milky Way. We focus in
particular on the effects of mass loss, dynamical friction and substructure--substructure interactions, the last of which has previously been ignored in analytic models of substructure
evolution. 
Our semi-analytical treatment reproduces both the spatial distribution of substructures and their mass function as obtained from the
most recent N-body cosmological calculations of Gao et al. (2004).
We find that, if mass loss is taken into account, the present distribution of substructures is practically insensitive to dynamical friction and scatterings from other substructures.
 Implementing these phenomena leads to a slight increase ($\simeq 5$\%) in the number of substructures at $r<0.35 r_{\rm vir}$, whereas their effects on the mass function are negligible. We find that mass loss processes lead to the disruption of substructures before dynamical friction and gravitational scattering can significantly alter their orbits.\\
Our results suggest that the present substructure distribution at $r>0.35 r_{\rm vir}$ reflects the orbital properties at infall and is, therefore, purely determined by the dark matter environment around the host halo and has not been altered by dynamical evolution. 
\end{abstract}

\section{Introduction}\label{sec:intro}

The cold dark matter paradigm for the formation of structure in the
Universe has met with considerable success in describing both the
large scale distribution of matter in our Universe (Bacon, Refregier
\& Ellis 2000, Percival et al. 2001, Spergel et al. 2003) and the
structure of galaxies and galaxy clusters (see for example Klypin,
Zhao \& Somerville, 2002, Kneib et al. 2003; although potential
discrepancies remain, see for example Weldrake, de Blok \& Walter
2003, Sand et al. 2004). In recent years, high resolution N-body
simulations have demonstrated that a record of the hierarchical
merging process through which dark matter haloes form in such
cosmologies is preserved in the form of substructure within haloes
(Moore et al. 1999, Klypin et al. 1999, Springel et al. 2001). When a
small halo merges into a large, ``host'', system it is not immediately
destroyed, but instead begins to orbit within the host, gradually
losing mass due to the actions of tidal forces.

These substructures are presumably the locations of galaxies in
clusters and of dwarf galaxies in the Local Group (Kauffmann, White \&
Guiderdoni 1993, Bullock, Kravtsov \& Weinberg 2000, Somerville 2002,
Benson et al. 2002b). The properties and evolution of substructures
have been extensively studied in recent years. The distribution of
halo masses and the distribution of substructures throughout the host
are now well studied (though that is not to say agreed upon) through
both N-body and semi-analytical means (Ghigna et al. 2000, de Lucia et
al. 2004, Gao et al. 2004, Taylor \& Babul 2004).

Although substantial computational efforts have been made to analyse the present distribution of substructures in dark matter
haloes, the subject still lacks detailed theoretical studies that explore the main processes that drive the evolution of those
systems.  In particular, one aspect of substructure evolution that has not been examined in detail is the interactions between
substructures. Although such interactions are self-consistently included in N-body simulations their importance has not been
assessed, while previous semi-analytical calculations have ignored interactions between substructures\footnote{Springel et al. (2001)
included mergers between sub-haloes in their combined N-body--semi-analytic study of a $z=0$ galaxy cluster, but did not assess in
detail the importance of these mergers. Somerville \& Primack (1999) also allowed for the possibility of mergers between
substructures in their semi-analytic model of galaxy formation. Note that neither of these studies addressed the questions we are
interested in here, namely the effects of substructure--substructure \emph{encounters} as opposed to mergers.}. These encounters are an
additional source of heating for substructures, thereby altering their rate of mass loss, and also change the distribution of
substructure orbits (and therefore their spatial distribution).

In this work we examine whether dynamical evolution occurring within the host halo alters the present distribution of dark
matter substructures.  We use a semi-analytic model that implements the main dynamical processes acting on substructures:
dynamical friction, gravitational scattering, gravitational heating and mass loss. Semi-analytic codes have several advantages
over N-body calculations: (i) they allow us to isolate the effects of each phenomena, (ii) they do not suffer from resolution
limitations and (iii) they require fewer computational resources. However, these codes are based on analytic approximations to
complex dynamical processes. This inevitably limits the accuracy of such codes. However, such models have been shown to provide a
good description for substructures in cosmological dark matter haloes (Taylor \& Babul 2004, Benson et al. 2002a). We will quantify
how the neglect of substructure-substructure interactions impacts upon the results of such models.

Ultimately, the question that we are attempting to answer in this work is the following: Is the present distribution of
substructures driven by physical processes acting in the host halo (in the sense that the orbits of those systems are strongly
altered by dynamical processes), or, alternatively, does their distribution reflect their orbital properties at the time of
accretion? The answer to that question is, for instance, of great importance for observational surveys of satellite galaxies,
which use these bodies as a direct indicator of the properties of the host haloes and the dark matter environment (e.g. Prada et
al. 2003, Sales \& Lambas 2004).
In that sense, knowing whether dynamical effects are important for the substructure distribution could help understand whether such effects can significantly alter the properties of observed satellite galaxies.

%{\bf AJB: I'm not sure what you mean here. Could you add some additional explanation?} 
%Additionally, this work may indirectly help
%to determine how the spatial and time resolution of N-body simulations may affect the distribution of substructures by taking into
%account that two-body interactions are highly sensitive to those quantities\footnote{For example, Prugniel \& Combes (1992) and
%Wahde \& Donner (1996) find that dynamical friction is artificially increased due to numerical noise if the particle number is
%small.}.

The remainder of this paper is arranged as follows. In \S\ref{sec:init_distrib} we describe how we generate host haloes and their
population of infalling substructures. In \S\ref{sec:sac} we describe the semi-analytic code used to evolve the substructures. In
\S\ref{sec:calc} we describe the calculations carried out in this work. We first show an illustrative study of the
substructure-substructure encounters that these systems suffer during their evolution. We then analyse the evolution of the
density profile and the orbit of substructures and examine how dynamical friction, gravitational scattering and mass loss affect
the present spatial and mass distribution of substructures. We also examine the evolution of the substructure mass
function. Lastly, in \S\ref{sec:concl} we summarise our results.

\section{Initial samples of dark matter substructures}\label{sec:init_distrib}

We generate initial conditions for our semi-analytical calculations by constructing merger trees for a sample of dark matter haloes
identified at $z=0$. Specifically, we begin with dark matter haloes of mass $2 \times 10^{12}h^{-1}M_\odot$, similar to that of the
Milky Way (Klypin, Zhao \& Somerville 2002). We then apply the methods of Cole et al. (2000) to construct the merging history of
this halo back to high redshift. For this calculation we assume a $\Lambda$CDM cosmology with parameters $\Omega_0=0.3$,
$\Lambda_0=0.7$, $\sigma_8=0.9$, $h(=H_0/100\hbox{km s}^{-1}\hbox{ Mpc}^{-1})=0.7$ and $\Gamma=0.21$, together with an
inflationary initial power spectrum (i.e. $P(k)\propto k^n$ with $n=1$). At each point in time, we identify the most massive
progenitor of the $z=0$ halo and classify this as the ``host'' halo at that redshift. We then identify haloes which are about to
merge with the host at each time and label these as new satellites.

At each point in time, we record the virial mass, virial velocity and concentration of the host halo (which is assumed to have an
NFW density profile). We also record the same information for each satellite along with the epoch of each merging event. We select
an initial orbit for each satellite according to the method of Benson (2004)---specifically, the initial position is chosen at
random on the surface of a sphere with radius equal to the current virial radius of the host halo, while the initial velocity is
chosen from the distribution found by Benson (2004) for $z=0$ in $\Lambda$CDM cosmologies\footnote{Benson (2004) finds some
evidence for evolution of the velocity distribution to higher redshifts, and for dependence on the masses of the merging
haloes. These dependencies are poorly quantified, and so we ignore them here.}.

In this way we generate both the formation history of the host halo
and the properties of the infalling population of satellites. These
are used as inputs to our model for the evolution of substructures.

\begin{figure}
\vspace{9.9cm} \includegraphics{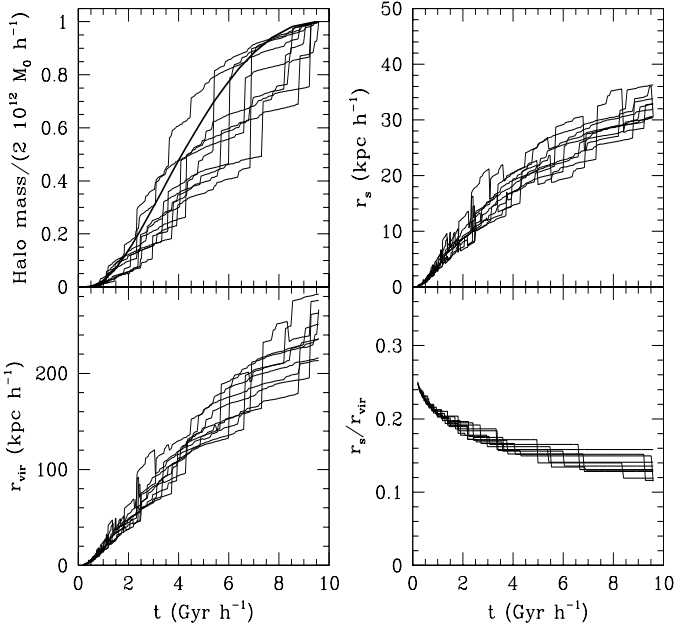}
\caption{Growth of the host halo in time. In each panel we plot the relevant property of each host halo from our sample of
nine. \emph{Upper left-hand panel:} The virial mass. Strong-full line shows the result of van den Bosch (2002) for a halo with $M_{\rm h}(z=0)=2\times 10^{12}h^{-1}M_\odot$ \emph{Upper right-hand panel:} The scale radius of the NFW density
profile. \emph{Lower left-hand panel:} The virial radius. \emph{Lower right-hand panel:} The concentration parameter. }
\label{fig:halo}
\end{figure}

\begin{figure}
\vspace{9.9cm} \includegraphics{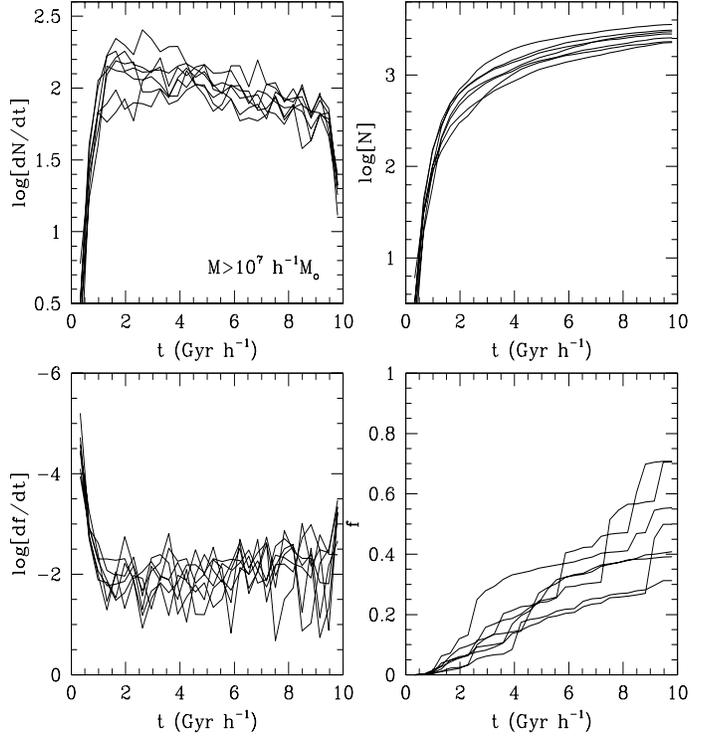}
\caption{\emph{Upper-left panel:} Evolution of the number of mergers per unit time. \emph{Upper-right panel:} Cumulative number of
mergers as a function of time. \emph{Lower-left panel:} Evolution of the mass fraction of accreted substructures (see text) per
unit time. \emph{Lower-right panel:} Cumulative mass fraction as a function of time. Note that only substructures with an initial
mass larger than $10^7 h^{-1} M_\odot$ were included in these calculations.}
\label{fig:merger}
\end{figure}

We generate a sample of nine such Milky Way-like haloes, which have $M_{\rm h}(z=0)=2\times 10^{12} h^{-1} M_\odot$, where $M_{\rm h}$ is the
total mass of the host halo, including the mass enclosed in dark matter substructures. In Fig.~\ref{fig:halo} we plot
the evolution of their mass (upper-left panel), virial (lower-left panel) and scale radii (upper-right panel) over one Hubble
time. As a comparison, the strong-solid line shows the averaged mass growth of a Milky Way-like halo obtained by van den Bosch (2002) using an extended Press-Schecter formalism.  In the lower-right panel, we also plot the variation of their concentration parameter (defined as the ratio between the
scale and virial radii) in time. It is interesting to note that both the mass and the virial radius of haloes grow almost
linearly with time on average until the present epoch.

For each realization in our sample, a different merger tree is defined. In Fig.~\ref{fig:merger} we show the number of
substructures merging with the host halo per unit time (upper-left panel) and the cumulative number of merging events during one
Hubble time (upper-right panel). Both panels show that the merging rate has slowed down since the early formation of the host
halo.

Each substructure has associated with it a virial mass. In the lower panels we plot the mass fraction $f$, which is defined as the
mass enclosed in substructures, normalised to $M_{\rm h}(z=0)$, which is accreted per unit time (lower-left panel) and the cumulative
mass fraction (lower-right panel) as a function of time. This last quantity accounts for the mass gain of the host halo as a
result of substructure mergers and, as we can see, ranges from 0.3--0.7 at $z=0$. Interestingly, despite the decrease in the
number of mergers, the cumulative mass fraction shows an approximately linear growth owing to the larger masses of the late-time
merging substructures.  For this plot, we consider only those substructures with $M\geq 10^7 h^{-1} M_\odot$. 
%{\bf JP: 
We note that the mass fraction enclosed in subhaloes barely depends on the arbitrary minimum mass selected here since $f$ is principally determined by a relatively low number of fairly massive substructures, in particular by those with $M (t_m)> 0.1 M_{\rm h}(z=0)\simeq 2\times 10^{11}h^{-1} M_\odot$ (see Section~\ref{sec:mass}), where $t_m$ is the merging time .
%}
% {\bf AJB: We should comment on the mass spectrum here --- i.e. if we resolved down to $10^6 h^{-1}M_\odot$ how would the fraction change?}

In order to obtain a better understanding of the complex processes acting on dark matter substructures during their evolution we
inject two control substructures in the initial sets with orbital and internal properties at their merging time $t=t_m$ selected
{\em ad hoc}. The scattering force induced by these bodies was settle to be zero for consistency. Control substructures are assumed to follow NFW profiles. For these galaxies, we shall perform an illustrative study
of the effects induced by substructure-substructure interactions.  In Table~\ref{tab:control} we show their orbital and structural
parameters.

\begin{table*}
\begin{tabular}{l l  l l l l } \hline \hline
& $M_{\rm h}$  & $r_{\rm s}$   & $r_{\rm vir}$ & $r(t_{\rm m})$  & $ e $\\ 
&$[M_\odot h^{-1}]$ & [kpc $h^{-1}]$ & [kpc $h^{-1}]$  & [kpc $h^{-1}]$ & \\
\hline
 Control A & $10^{8} $ & 0.90  &  6.24  & 133.90 &0.00\\
 Control B & $10^{8} $ & 0.90  &  6.24  & 133.90 &0.50\\
\hline \hline
\newline
\newline
\label{tab:calc}
\end{tabular}
\caption{Control substructures. Orbital and density parameters at $t=t_{\rm m}$. $e$ denotes the orbital eccentricity, defined as
$e=(r_{\rm a}-r_{\rm p})/(r_{\rm a}+r_{\rm p})$, where $r_{\rm a}$ and $r_{\rm p}$ are the orbital apo and peri-centres, respectively. Control substructures are
initially located at the apo-centre, $r(t_{\rm m})$. }
\label{tab:control}
\end{table*}

\section{The semi-analytical code}\label{sec:sac}

In this section we present a description of the semi-analytical code
used to model the evolution of the dark matter substructures
described above as they orbit in the host potential. 

%This code has
%several advantages over N-body simulations: (i) it does not suffer
%from resolution limitations, (ii) it permits switching on and off of
%the main mechanisms that drive the evolution of substructures in
%order to individually assess their effects and (iii) it is several
%orders of magnitude faster. However, it is based on analytic
%approximations of complex dynamical processes. This inevitably limits
%the scenarios in which it can reasonably be used, but has been shown
%to provide a good description for substructures in cosmological dark
%matter haloes (Taylor \& Babul 2004, Benson et al. 2002a)

\subsection{Dynamics of dark matter substructures}

We treat dark matter substructures as point-mass particles that move within a host halo potential in order to compute their
unperturbed orbits. Additionally, we include a dynamical friction force and account for forces between pairs of dark matter
substructures by direct summation.

Our model solves the equation of motion:
\begin{eqnarray}\label{eqn:f_total}
\ddot{\bf r}={\bf f}_{\rm h} + {\bf f}_{\rm sub} +{\bf f}_{\rm df},
\end{eqnarray}
where ${\bf f}_{\rm h}$ is force per unit mass (or {\em specific
force}) induced by the host halo, ${\bf f}_{\rm sub}$ is that due to
interacting dark matter substructures and ${\bf f}_{\rm df}$ is the
dynamical friction force exerted by the diffuse background of dark
matter particles on the substructure.

We assume that the host halo, as well as the dark matter
substructures, follow an NFW density profile (Navarro, Frenk \& White
1997)
 
\begin{eqnarray}\label{eqn:nfw}
\rho_{\rm NFW}=\frac{\rho_{0}}{(r/r_{\rm s})(1+r/r_{\rm s})^2},
\end{eqnarray}
where $\rho_0$ is a constant and $r_{\rm s}$ is the halo scale length. The force, ${\bf f}_{\rm h}$, acting on a particle at the
position ${\bf r}$ can be analytically derived
 \begin{eqnarray}\label{eqn:f_h}
{\bf f}_{\rm h}=-\frac{\G M_{\rm h}}{r^2} \frac{ \ln [1 + r/r_{\rm s}] -r/(r+r_{\rm s})}{\ln [1 + r_{vir}/r_{\rm s}] -r_{\rm vir}/(r_{\rm vir}+r_{\rm s})}\hat {\bf r},
\end{eqnarray}
where $M_{\rm h}$ is the mass within the virial radius $r_{\rm vir}$ after subtracting the mass bound in substructures, and $\hat {\bf
r}\equiv {\bf r}/r$. We note that $M_{\rm h}, r_{\rm s}$ and $r_{\rm vir}$ are time-dependent quantities (see Fig.~\ref{fig:halo}).

The presence of $N_{\rm sub}$ substructures will perturb the orbits that dark matter substructures would follow in a smooth
potential. To estimate the perturbation we perform a simple force summation, so that the specific force acting on the $i^{\rm th}$
substructure is
 \begin{eqnarray}\label{eqn:f_sub}
{\bf f}_{i,\rm sub}=\Sigma^{N_{\rm sub}}_{i\neq j}\G M_j\frac{{\bf r}_j-{\bf r}_i}{(\epsilon_i^2+ | {\bf r}_j-{\bf r}_i  |^2)^{3/2}}
\end{eqnarray}
where $\epsilon_i$ is the {\em softening parameter} of the $i^{\em th}$ substructure which accounts for its finite extent. Our approximation of point masses is valid providing encounters remain in the regime of $| {\bf r}_j-{\bf r}_i | >
r_{i,{\rm vir}}+ r_{j,{\rm vir}}$. Closer encounters, where two systems overlap, can be treated as inelastic collisions. In such
cases, a large fraction of the orbital energy is transmitted to stars through resonances so that eq.~(\ref{eqn:f_sub})
overestimates the resulting velocity change.  In order to compare our results to those of self-consistent N-body simulations we
have chosen $\epsilon_i\equiv r_{i,{\rm h}}$, $r_{i,{\rm h}}$ being the half-mass radius. This choice decreases the relative force
for overlapping encounters and avoids a diverging force at very small relative distances, whereas if $| {\bf r}_j-{\bf r}_i |>
r_{i,{\rm vir}}+ r_{j,{\rm vir}}\sim 2 r_{i,{\rm vir}} $ it recovers the Newtonian behaviour.

%\begin{figure}
%\vspace{9.0cm} \special{psfile=eps.ps angle=0 hscale=46
%vscale=46 hoffset=-20 voffset=-65}
%\caption{ Normalised
%relative forces as a function of distance for two sets of
%substructure parameters: (i) solid and dotted lines $(r_{\rm
%s},r_{\rm h},r_{\rm vir})=(0.7,1.6,3.5)$ kpc and (ii) dashed and
%dotted-dashed lines $(r_{\rm s},r_{\rm h},r_{\rm vir})=(1.4,2.5,5.3)$
%kpc. Dotted and dotted-dashed lines are NFW forces (eq.~\ref{eqn:f_h})
%whereas solid and dashed reproduce softened point-mass forces
%(eq.~\ref{eqn:f_sub}).}
%\label{fig:eps}
%\end{figure}

Several studies of satellite orbital decay have shown that, in spherical systems, Chandrasekhar's formula for dynamical friction
(Chandrasekhar 1943) is sufficiently accurate if the Coulomb logarithm is treated as a free parameter to fit to N-body data
(e.g. van den Bosch et al. 1999, Colpi et al. 1999). Semi-analytic methods that include Chandrasekhar's dynamical friction have
been demonstrated to reproduce accurately the overall evolution of satellite galaxies (e.g. Vel\'azquez \& White 1999, Taylor \&
Babul 2001, Benson et al. 2004) and, therefore, represent a useful tool for conducting extensive studies of a large parameter
space. Chandrasekhar's formula can be written as
\begin{equation}
{\bf f}_{\rm df}=-4\pi \G M \rho(<v) {\rm ln}\Lambda \frac{{\bf v}}{|{\bf v}|^3}  ~,
\label{eqn:df}
\end{equation}
where ${\rm ln}\Lambda$ is the Coulomb logarithm and
\begin{equation}
\rho(<v)=\rho_{\rm NFW}(r)\left[{\rm erf}(X)-\frac{2 X}{{\sqrt \pi}}{\rm e}^{-X^2}\right],
\label{eqn:ff}
\end{equation}
where $X=|{\bf v}|/{\sqrt 2} \sigma_{\rm h}$ and $\sigma_{\rm h}$ is the one-dimensional velocity dispersion of the halo dark
matter,
defined as $\sigma\equiv1/\rho_{\rm NFW}(r)\int_\infty^r\rho_{\rm NFW}(r')f_h(r') dr'$ (e.g., eq.~2.12 of Hernquist 1993).
%{\bf AJB: We should state what value we use for $\sigma_{\rm d}$.} 
We fix $\ln \Lambda=2.1$ from the results of
Pe\~narrubia, Just \& Kroupa (2004).

We use a Runge-Kutta algorithm to solve eq.~(\ref{eqn:f_total}) with a
time-step $\Delta t=6.5\times 10^{-4} h^{-1}$ Gyr, short enough to properly
resolve close encounters. Subsequently, we apply a leap-frog
integration of the orbit at each time-step.

\subsection{Mass evolution}

Substructures experience mass loss whenever the external tidal field is sufficiently strong. Mass which becomes unbound can
subsequently escape from the satellite. Depending on the rate of change of the external tidal field as seen along the satellite
orbit, we can distinguish between slow and rapid regimes: tidal limitation and tidal heating, respectively.

\subsubsection{Tidal limitation}

In the first regime, the unbound mass is determined by the tidal radius (King 1962), which is defined for a spherically symmetric
satellite as the distance from the satellite centre at which the force due to the satellite's self-gravity and the external tidal
force balance. If the satellite follows a circular orbit, the system can be considered as static in a rotating frame, and one can
estimate the tidal radius as (King 1962)
\begin{equation}
R_{\rm t}\approx \bigg(\frac{G M_{\rm s}}{\omega^2-\d^2\Phi_{\rm h}/\d
r^2}\bigg)^{1/3},
\label{eqn:Rt}
\end{equation}
where $\omega$ is the angular velocity of the satellite and $\Phi_{\rm
h}$ the potential of the host halo.

This approximation is valid when: (i) The substructure mass is much smaller than that of the host halo and 
(ii) $R_{\rm t}$ is small compared to the distance to the centre of the host halo ($R_{\rm t}\ll
r$). Even under these conditions, the mass within $R_{\rm t}$ is not exactly equal to the bound mass, since there may be particles
that have (small) positive energy and stay in transient orbits within the satellite (e.g. Binney \& Tremaine 1987). Furthermore,
some particles at smaller radii, which nevertheless have orbits which take them to larger radii, can become unbound (Kampakoglou
\& Benson, in preparation).

If the satellite follows a non-circular orbit one can still use
eq.~(\ref{eqn:Rt}) to calculate the instantaneous tidal radius, with
$\omega$ now being the instantaneous angular velocity. For these
orbits, the mass loss is most rapid near peri-centre where the tidal
force is greatest (e.g., Pe{\~n}arrubia, Kroupa \& Boily 2002, Piatek
\& Pryor 1995). In our calculations, the effects of transient orbits
are assumed to be negligible compared to the total amount of mass
stripped out by the peri-galacticon passages.

Whereas eq.~(\ref{eqn:Rt}) is quite accurate in accounting for the
mass loss induced by tidal forces from the host halo, close
encounters between dark matter clumps may enhance the mass stripping
process by heating (i.e. expanding) these systems. Such collisions
occur on time-scales short compared to the orbital period and can be
treated as tidal shocks.

\subsubsection{Tidal heating}

Satellites travelling through regions where the external potential
changes rapidly suffer tidal shocks. This process can be modelled as a
perturbation, with a particular frequency spectrum, that adds energy
to the satellite particles.

Gnedin \& Ostriker (1999) show that, as a result of the shock, the
satellite is contracted, with subsequent expansions and
re-contractions until it reaches a final state of equilibrium, in
which the binding energy is reduced compared to the original value and
the satellite has expanded. This non-equilibrium phase lasts, in their
models, for around 20 satellite dynamical times.

The detailed study of tidal shocks is beyond the scope of this
work. We choose to follow the method of Gnedin, Hernquist \& Ostriker
(1999), hereinafter GHO, in order to calculate the energy change of
the substructure.

In the impulse approximation, the mean change in the specific energies
of particles in a spherical system, $i$, due to the encounter with an extended body,
$j$, can be written as
 \begin{eqnarray}\label{eqn:eb}
\left< \Delta E_1\right>_i=\left<\frac{1}{2}(\Delta
v)^2\right>_i=\frac{4}{3}\bigg( \frac{G M_j}{V_0 b^2}\bigg)^2
R^2 \chi(R)
\end{eqnarray}  
where $b,V_0$ are the relative distance and velocity
between two substructures at the moment of closest approach, $M_j$ is
the mass of the system $j$ and $R$ is the particle's radius in the
substructure. There are several corrections to this equation which
must be taken into account:
\begin{enumerate}
\item {\bf Extended bodies}: The function $\chi(R)$ accounts for the
fact that the energy change of particles belonging to the system $i$
depends on the tidal field of the body $j$ through its density
profile. For the sake of brevity, we refer the reader to eq. (11)-(14)
and eq. (17) of GHO for a description of the expression $\chi(R)$.
\item {\bf Diffusion term}: Gnedin \& Ostriker (1999) show that the
second-order energy gain $\langle\Delta E^2\rangle=\langle ({\bf v}
\Delta{\bf v})^2\rangle$ induces a shell expansion stronger than
$\langle\Delta E\rangle$. This term can be implemented in our
semi-analytic calculations as a multiplicative correction $\lambda$ to
the first-order energy variation. Here we implement the results of
Taylor \& Babul (2001) who, using a similar mass-loss scheme, estimate
$\lambda=2.8$ from fitting the mass evolution to N-body simulations.
\item {\bf Adiabatic correction}: In the impulse approximation, one assumes that particles do not move during the
encounter. However, in reality particles in the substructure \emph{will} move during the encounter, leading to a suppression of the
heating. eq.~(\ref{eqn:eb}) can be corrected for this fact by including an adiabatic correction term
\begin{eqnarray}\label{eqn:ac}
A(x)=(1+x^2)^{-\gamma}
\end{eqnarray}   
where $x\equiv t_{\rm sh}/t_{\rm orb}(R)$, i.e., the ratio between the shock duration and the substructure's internal orbital
period at a given radius. Specifically
\begin{eqnarray}\label{eqn:tsh}
t_{\rm sh} & = & \frac{b}{V_0} \\ \nonumber
t_{\rm orb}(R) & = &2\pi\frac{R}{v(R)}.
\end{eqnarray}   
and $v(R)$ is the rotation curve of the substructure $i$. The exponent $\gamma$ depends on the encounter duration relative to the
half-mass dynamical time, $t_{\rm dyn}$ of the system $i$, and varies from $\gamma=2.5$ for $t_{\rm sh}\le t_{\rm dyn}$ to
$\gamma=1.5$ for $t_{\rm sh} \ge t_{\rm dyn}$ (Gnedin \& Ostriker 1999). The adiabatic correction tends to unity for $x\ll 1$
(typical of the outer parts of substructures) and to zero for $x\gg 1$ (typical of the central parts).

\item {\bf Time-varying correction}: Eq.~(\ref{eqn:eb}) uses the maximum value of the tidal field (that is, the tidal field at the
moment of closest approach between both substructures) in order to calculate the particles' energy change. However, in reality,
the interaction occurs along the whole encounter. Gnedin \& Ostriker (1999) show that the external tidal force can be approximated
by a Gaussian centred at the moment of closest approach and with dispersion equal to the encounter duration, obtaining a
correction to eq.~(\ref{eqn:eb}) from fits to N-body simulations that can be written as
\begin{eqnarray}\label{eqn:t_corr}
I_{\rm corr}(y)=(1+y^2)^{-3/2},
\end{eqnarray} 
where $y = t_{\rm sh}/t_{\rm dyn}$.
\end{enumerate}

After combining eq.~(\ref{eqn:eb}) with corrections (ii) through (iv),
the resulting equation for the tidal heating induced by close
encounters between substructures is

\begin{eqnarray}\label{eqn:e_sh}
\Delta E=\Delta E_1 \lambda A(x) I_{\rm corr}(y).
\end{eqnarray} 

To determine the changes experienced by the satellite after the energy injection, we assume that no shell crossing occurs. This
approximation is quite accurate since the encounters between substructures lead to small changes of the internal energy due to
their low masses and in this limit the change in radius of each shell, $\Delta R$, is a rapidly increasing function of radius (see
below). Under this condition, a change of energy $E\rightarrow E+\Delta E$ results in an expansion of the system $i$
\begin{equation}
\Delta R=\frac{\Delta E R^2}{G M_i(R)},
\label{eqn:deltar}
\end{equation}   
so that the mass distribution after the tidal shock is
$M^\prime_i(R^\prime)=M_{\rm i}^\prime(R+\Delta R)=M_{\rm i}(R)$. As
this equation suggests, some material will be expanded out of the
tidal radius which will, therefore, enhance the degree of mass
loss. Additionally, the mass loss induces a progressive shrink of the tidal radius since $R_t\propto M^{1/3}$.
We note that the shell radii expand monotonically in time after each tidal
shock.

Our code computes the mass of substructure $i$ at each time-step in the following way:\\ 1) Calculate $R_t$ using
eq.~(\ref{eqn:Rt}).\\ 2) If the distance between the substructure and any other substructure reaches a minimum during the
time-step, the code computes $\Delta E$ and $\Delta R$ for each $R$ in the substructure and subsequently computes the expansion of
the substructure.\\ 3) Finally, the code calculates the new mass of the satellite $M^\prime=M(R<R_t)$. If $M' > M$ then we force
$M'=M$, so that the mass of the substructure decreases monotonically.

As Taylor \& Babul (2001) note, this technique suffers from some limitations: (i) The assumption of virial equilibrium between the
satellite shocks.\footnote{The estimation of the mass shell expansion applied in this work assumes that an extended system virializes after a tidal shock event. Gnedin \& Ostriker (1999) show that the viralization process lasts approximately 20 dynamical times of the system. Thus, our method may not consistently reproduce the subhalo heating if the system suffers strong encounters within short time intervals (fortunately, we shall show in Section~\protect\ref{sec:int} that such interactions are fairly rare).}
 (ii) Numerical simulations of tidal mass stripping have shown that the process is highly anisotropic. Our mass loss scheme does not account for any possible net angular
momentum of the escaping particles. 
(iii) Overlapping encounters cannot be accounted for by this simple mass-loss scheme. In such
cases, depending on the mass ratio of the clumps and their relative velocity, the encounter may result in the destruction of one
or both substructures. Even taking account of these limitations, this scheme has been proven to lead to accurate estimates of the
mass evolution (Taylor \& Babul 2001, Pe{\~n}arrubia 2003).

\section{Results}
\label{sec:calc}

The goal of this paper is to develop a better understanding of the dynamical processes that may influence the present distribution
of dark matter substructures. The main advantage of the semi-analytic code outlined above is that we can study separately the
effects induced by each process.

In Table~\ref{tab:sim} we describe the calculations carried out for this work. Simulations C1,C2,C3 and C4 combine the
implementation of dynamical friction and gravitational scattering in the semi-analytic code. The calculation C5 switches off the
mass evolution scheme by imposing a constant mass for all substructures during their evolution. Lastly, simulation C6 repeats the
calculation of C1, but with a minimum mass at merging for substructures which is ten times larger than for the other simulations.

For each of the calculations, we evolve the substructure samples described in Section~\ref{sec:init_distrib}.
\begin{table*}
\begin{tabular}{||l |l  |l |l |l ||} \hline \hline
Name& mass resol. ($m$) & mass loss   & dyn. friction & scattering  \\ 
& [$h^{-1}M_\odot$] &&&\\\hline
 C1& $10^{7}~ (m_2$) & yes  &   yes & yes \\
 C2& $10^{7}~ (m_2$) & yes  &   no  & yes \\
 C3& $10^{7}~ (m_2$) & yes  &   yes & no \\
 C4& $10^{7}~ (m_2$) & yes  &   no  & no \\
 C5& $10^{7}~ (m_2$) & no   &   yes & yes \\
 C6& $10^{8}~ (m_1$) & yes  &   yes & yes \\
\hline \hline
\newline
\newline
\end{tabular}
\caption{Calculations carried out. We study different mass resolutions by evolving only those substructures that at the time of merging have $M>m$. We also analyse the effects that dynamical friction and gravitational scattering induce on the evolution of substructures. We note that in those calculations without substructure scattering, the tidal heating scheme was also switched off for consistency.}
\label{tab:sim}
\end{table*}

\subsection{Interactions between dark matter substructures}\label{sec:int}
The internal evolution of dark-matter substructures might be affected by repeated encounters with other substructures. In contrast
with the effects of two-body collisions with particles of the {\em smooth} dark matter component (which can be treated
analytically as dynamical friction), substructure-substructure encounters may lead to a strong tidal heating (i.e. to an overall
expansion of the system) which would accelerate mass stripping due to tidal forces. That physical process, however, will induce significant effects only when the encounter is sufficiently close.

In order to illustrate the properties of two-body encounters between substructures within a given host halo, we use two
substructures (control A and B, see Table~\ref{tab:control}) injected in each of our host halo samples with kinematical and
internal properties selected {\em ad hoc}. The full dynamical scheme (C1) was selected. For each of those substructures, we
calculate at each time step the relative distance to all other substructures. If a minimum occurs, we consider that an encounter
takes place and record the orbital parameters. 

\begin{figure}
\vspace{10.5cm} \includegraphics{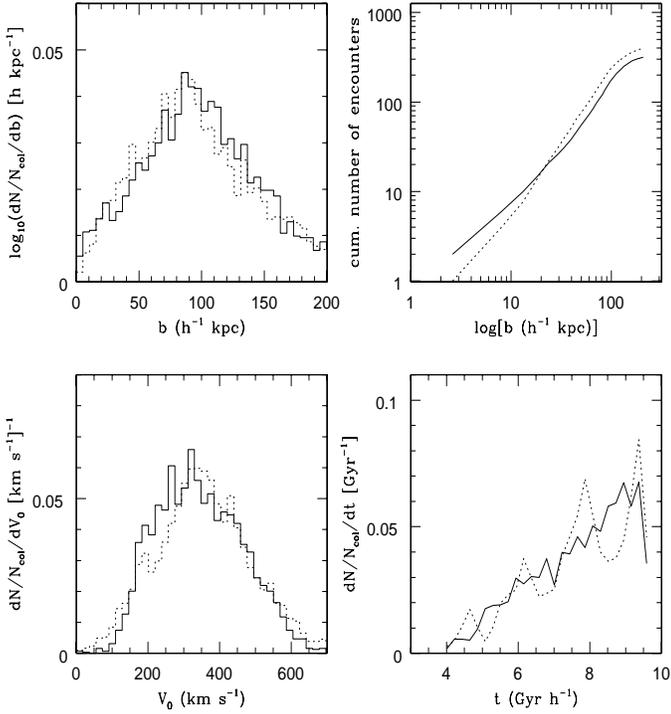}
\caption{ Distributions of minimum approach distances
(upper-left panel) and relative velocities at the moment of
minimum approach (bottom-left panel) for the control galaxies A  (solid lines) and B (dotted lines) of Table~\ref{tab:control}.  In the upper-right panel we also plot the cumulative number of encounters as a function of minimum distance. In the bottom-right panel we show the encounter rate as a function of time. }
\label{fig:imp}
\end{figure}

 In Fig.~\ref{fig:imp} we plot the distribution of encounter parameters for all
encounters that our control substructures experience during the last 5.5 $h^{-1}$Gyr in all host haloes. In the upper-left panel we show
the distribution of minimum encounter distances for our control substructure A (circular orbit) and B (eccentric orbit,
$e=0.5)$. Both substructures show curves with a marked maximum at $b\simeq 85 h^{-1}$ kpc (comparable to the size of the
haloes). For both orbits, the number of close encounters $b \approx 0$ is fairly small, as can be seen from the upper-right
panel, so that during the last 7 Gyr the average number of overlapping encounters (i.e. those with minimum encounter distances of
$b\simless 2 r_{\rm vir}\simeq 10 h^{-1}$ kpc, where $r_{\rm vir}$ is the substructure's virial radius) was less than 10,
independently of their initial eccentricity. This result favours our semi-analytic calculation of substructure orbits since
relative forces along non-overlapping collisions can be accurately reproduced by interactions of point-masses.  It is also interesting to note that the substructure following an eccentric orbit has suffered around 33\% more encounters than that following a circular one.
The distribution
of relative velocities at the moment of closest approach also shows unique maxima at $V_0\simeq 320$ km/s (control A) and
$V_0\simeq 340$ km/s (control B).  Lastly, in the lower-right panel we plot the evolution of the encounter rate ${\rm d}N/{\rm
d}t$ for both control substructures. Control A shows a linear growth of ${\rm d}N/{\rm d}t$, reflecting a progressive increase in
the number of substructures within the host galaxy (see Fig.~\ref{fig:merger}). The oscillations in the encounter rate of Control
B are due to its eccentric orbit and show that the number of encounters is enhanced at the peri-centre of the orbit. As for
control A, the mean rate within one orbital period grows linearly in time.

Although the results obtained in this section are purely illustrative, we may foresee that overlapping collisions are rare,
representing less than 2\% of the total number of collisions for our control substructures. More distant encounters will induce
negligible tidal heating since, from eq.~(\ref{eqn:eb}), we have that the heating energy scales as $\Delta E_1\propto 1/b^4$.

\subsection{Evolution of dark matter substructures}\label{sec:results}
In this section we analyse the effects that substructure-substructure interactions induce on the final distribution of
substructures within the host halo.

We note that, in order to compare our results to those of recent N-body calculations, our statistical study was carried out in a
way similar to that used when analysing cosmological N-body simulations: (i) each halo realization was evolved separately; (ii)
the resulting substructure samples were compiled into a single sample; (iii) we only consider substructures within the virial radius of
each of the host haloes and with final mass larger than the resolution mass (which we introduce artificially into our
calculation for the purposes of this comparison).

\subsubsection{Structural evolution }\label{sec:heat}
\begin{figure}
\vspace{10.5cm} \includegraphics{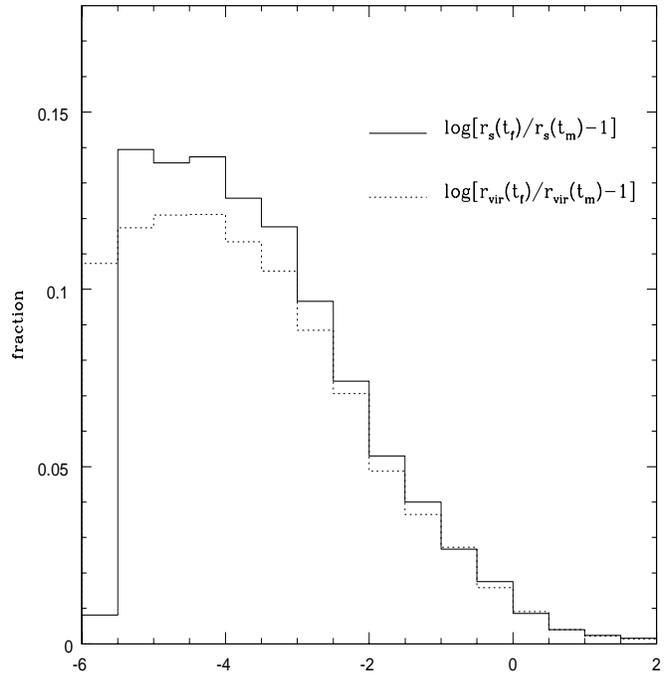}
\caption{Fraction of substructures as a function of their expansion ratio. The full and the dotted lines show the expansion ratio of the scale-length ($r_s$) and the virial radius ($r_{\rm vir}$), respectively. } 
\label{fig:rcrt}
\end{figure}

Following the above prescriptions, we determine, for the calculation C1, the average substructure expansion (heating) as a
consequence of substructure scattering. In Fig.~\ref{fig:rcrt} we show the fraction of substructures as a function of the
expansion ratio of the scale-length (full line) and the virial radius (dotted line). As can be seen, the fraction of substructures
that were inflated by more than 10\% of their initial size ($\log_{10}[r(t_f)/r(t_{\rm m})-1]\ge -1)$ due to tidal heating corresponds to
approximately 2\% of the total number of substructures. The remaining substructures suffer an expansion that can be considered
negligible for mass loss computations.

In light of this result, we expect tidal heating to induce negligible effects on the final substructure mass function.

\subsubsection{Orbital evolution}\label{sec:decay}
In this work we have assumed that the interaction of substructures with the rest of the dark matter present in the host halo can
 be separately treated as: (i) dynamical friction (smooth dark matter component) and (ii) substructure-substructure scattering
 (clumpy dark matter component). We have analysed the effects that each component induces on the orbits of substructures by
 switching on and off these terms in eq.~(\ref{eqn:f_total}). Due to the growth in mass and size of the host halo, adiabatic
 invariants (e.g see Section~3.6 of Binney \& Tremaine 1986 for details) are adequate quantities to reflect how those processes
 alter the orbits of substructures since they remain constant during adiabatic (slow) changes of the host halo potential. Thus,
 in our calculations, their values can only be altered either by dynamical friction or by substructure-substructure interactions.

\begin{figure}
\vspace{10.5cm} \includegraphics{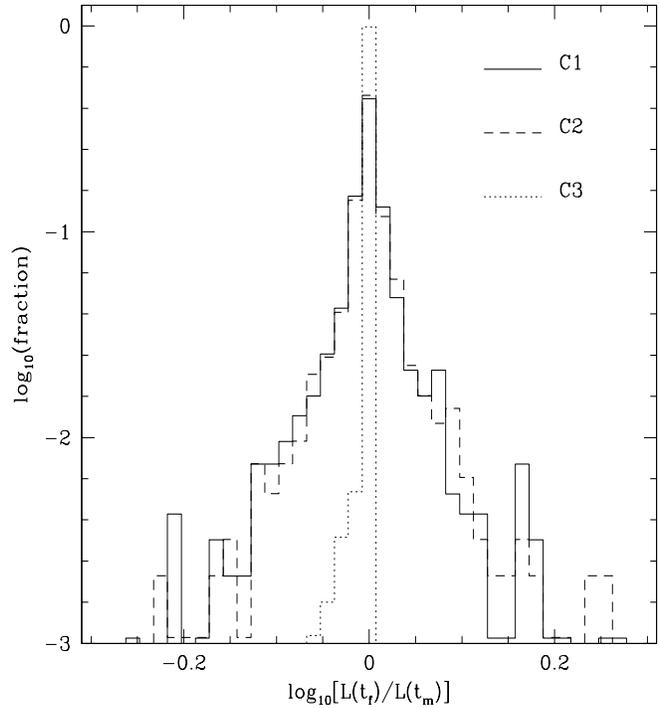}
\caption{Fraction of substructures as a function of the angular momentum variation. We show the distribution resulting from calculations where (i) the force term accounts for dynamical friction and substructure scattering (full line), (ii) the dynamical friction term is switched off (dashed line) and (iii) the substructure-substructure scattering term is switched off (dotted line). Substructures included in this sample have $M>10^7h^{-1}M_\odot$ and are located at $r<r_{\rm vir}$ at $z=0$.} 
\label{fig:angmom}
\end{figure}

The total angular momentum of systems orbiting in isotropic potentials is an adiabatic invariant. If only dynamical friction is
implemented, the angular momentum decreases as a consequence of the drag force, whereas if only scattering is implemented
substructures can gain or lose angular momentum depending on the characteristics of the two-body encounters occurring during their
evolution. In Fig.~\ref{fig:angmom} we plot the fraction of substructures as a function of their angular momentum variation. The
dotted line, calculation C2 (i.e. switching on dynamical friction) shows that the effect of the smooth dark matter component is
practically negligible, since fewer than 1\% of the surviving substructures have suffered a measurable angular momentum loss. This
result is fully in agreement with that of Zhao (2004,) who, on the basis of semi-analytical calculations of satellite orbits in a
growing host potential, conclude that the effects of dynamical friction are suppressed by mass loss. On the contrary, the effects
of substructure-substructure interactions (calculation C3; dashed line) on individual orbits of dark matter clumps are
considerably stronger. The distribution of the angular momentum variation follows approximately a Gaussian centred at
$L(t_f)/L(t_{\rm m})\simeq 1$ with dispersion of around 10\%. This figure shows that substructure scattering has altered the orbits of
the present substructures much more effectively than dynamical friction.  Implementing both dynamical friction and scattering
(calculation C1) produces basically no further change in the angular momentum distribution.
  
\begin{figure}
\vspace{10.5cm} \includegraphics{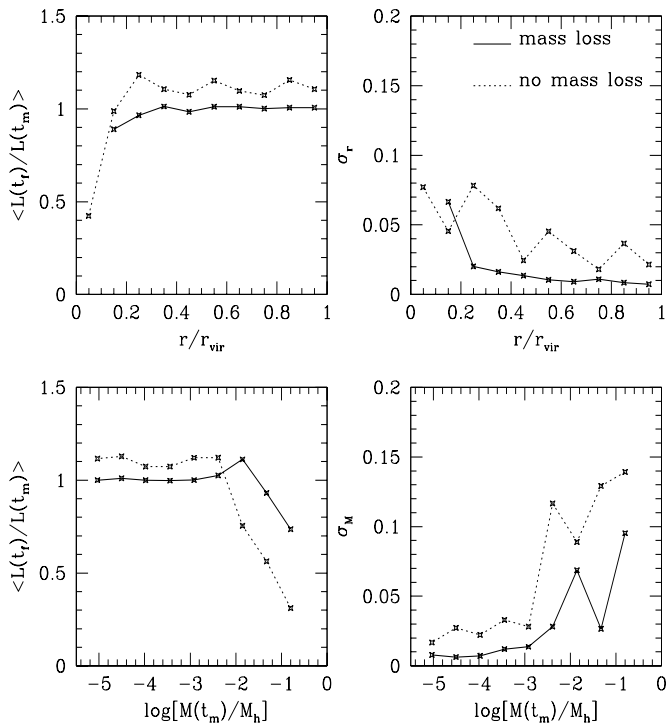}
\caption{Averaged change of angular momentum as a function of radius (upper-left panel) and initial mass (lower-left panel). We also plot the variance of the distributions as a function of radius (upper-right panel, denoted as $\sigma_r$) and initial mass (lower-right panel, denoted as $\sigma_M$). We compare calculations where mass loss is switched on (full line; C1) and off (dotted line; C5).} 
\label{fig:dell}
\end{figure}

On the basis of these results, the following question arises: Why are dynamical friction and gravitational scattering inefficient
at altering substructure orbits? To answer that question we compare in Fig.~\ref{fig:dell} the averaged angular momentum change
$\langle L(t_f)/L(t_{\rm m})\rangle$ and the deviation from the average $\sigma$ (i.e. the variance of the distribution) of
substructures evolved in calculations where our mass loss scheme has been switched on (full line; calculation C1) and off (dotted
line; calculation C5). We plot those quantities as a function of the substructure initial mass (i.e. at the merging time) and the
present orbital radius.

Looking first at the dotted lines, we observe that if we allow substructures to evolve until $z=0$ (i.e. they are not disrupted by
tidal fields), the averaged angular momentum change is considerable, independent of initial mass and present radius. Moreover, the
angular momentum variation as a function of mass (lower-left panel) resembles that typically found in stellar systems with a low
number of particles and a given mass spectrum where, owing to two-body relaxation processes, the angular momentum of most massive bodies decreases on average (see, e.g., Section 8.4 of Binney \&
Tremaine 1986). This panel shows that subhaloes with $M(t_{\rm m})>0.01 M_{\rm h}(z=0)$ lose angular momentum ($\langle L[t_f]/L[t_{\rm m}]\rangle<1$) on average whereas light subhaloes $M(t_{\rm m})<0.01 M_{\rm h}(z=0)$ gain
it ($\langle L[t_f]/L[t_{\rm m}]\rangle>1$). 
%The distribution that we observe in that panel is similar to that resulting from an {\emangular momentum equipartition}
The lower-right panel also shows that the deviation from the average ($\sigma$) is also larger for substructures that have lost a large fraction of their initial angular momentum.\\
Besides the dependence of the angular momentum distribution on the initial mass of substructures we also find a radial dependence. In the upper-left panel we show the averaged angular momentum variation as a function of radius. The dotted line shows that
substructures located at $r<0.12 r_{\rm vir}$ lose on average approximately 60\% of their initial angular momentum, whereas the angular
momentum of substructures at $r>0.35 r_{\rm vir}$ has increased by 10--20\%. From these results, it appears that 
changes in angular momentum and orbital radius are related, with massive substructures decaying to the inner regions of the host
halo, while light substructures are ejected to the outer regions as a result of two-body interactions.

The present angular momentum distribution is strongly modified if we allow substructures to lose mass (full line). In
that case, $\langle L(t_f)/L(t_{\rm m})\rangle\simeq 1$, independently of mass and radius. Only the most massive substructures
($M[t_{\rm m}]>0.05 M_{\rm h}[z=0]\simeq 10^{11}h^{-1}M_\odot$) show a decrease of angular momentum (up to 28\%). Substructures at $r<0.2 r_{\rm vir}$ also present some degree of dynamical evolution, with a distribution centred at $\langle
L(t_f)/L(t_{\rm m})\rangle\simeq 1$ but with a considerably larger deviation from the mean value $\sigma_r\simeq 0.07$ (in contrast with
those substructures at $r>0.2 r_{\rm vir}$, which show $\sigma_r\simeq$ 0.01--0.02).

These results appear to indicate that the mass loss process prevents significant alterations of the initial adiabatic invariants, thus minimising dynamical evolution effects such as orbital decay and gravitational scattering. Only massive subhaloes ($M[t_{\rm m}]>0.05 M_{\rm h}[z=0]$) show a measurable orbital evolution, but these represent less than 2\% of the total number of substructures in our sample. 

\subsection{Present distribution of substructures}\label{sec:dist}

\subsubsection{Spatial distribution}
In this Section we analyse how dynamical friction and substructure scattering affect the final radial distribution of
substructures. 
In order to compare our results with those from cosmological calculations at $z=0$, we mimic the criteria used to
identifying substructures: our subhalo sample includes only those substructures with masses above a certain value (i.e {\em mass resolution}). In N-body calculations, that threshold is determined by the number of particles  and the mass of the host system since only those systems with a minimum number of particles (equivalently, mass) can be distinguished as bound substructures at any time of the simulation. 
Additionally, we investigate the effects of the mass resolution by performing calculations with two different minimum masses (similar to those of Gao et al 2004).

\begin{figure}
\vspace{10.5cm} \includegraphics{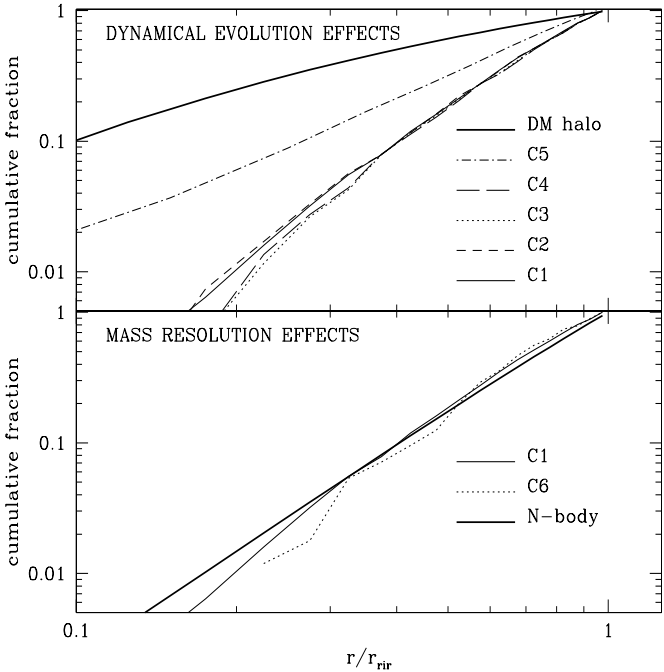}
\caption{Cumulative radial distribution of substructures at $z=0$. Upper panel presents the distribution resulting from using different force terms and that obtained if the mass loss treatment is switched off (see Table~\ref{tab:calc} for details). We also show for comparison the cumulative profile of the host halo. We note that those substructures included in the distribution have $M\ge10^7 h^{-1}M_\odot$ at $z=0$ and are located within the virial radius.
In  the lower-left panel we compare the cumulative profile that results from simulations where the minimum mass of substructures at their merging time was above $10^8 h^{-1}M_\odot$ (dotted line) and $10^7 h^{-1}M_\odot$ (full line).  
The strong full line shows the results of Gao et al. (2004) from large N-body simulations.}
\label{fig:r}
\end{figure}

In the upper panel of
Fig.~\ref{fig:r} we plot the present radial distribution of substructures for calculations where the different force terms of eq.~(\ref{eqn:f_total}) where switched on and off (see Table~\ref{tab:calc} for details). Additionally, the dot-dashed line shows the resulting distribution of sustructures if the mass loss scheme is not implemented and the heavy solid line shows the mass distribution of the host halo.  This panel shows that:\\ (i) The cumulative profile of
substructures is on all scales less concentrated than that of the smooth dark matter component.\\ (ii) The largest difference in
the final distribution is induced by mass loss. Switching on our mass loss treatment drastically reduces the number of
substructures on all scales. \\ 
(iii) The dynamical friction and scattering terms of eq.~(\ref{eqn:f_total}) only introduce differences of around 5\% in the number of subhaloes at $r\le 0.35 r_{\rm vir}$.\\
(iv) Our
semi-analytical treatment allows us to disentangle the contribution of dynamical friction and that of scattering. A comparison between the calculations C2 (scattering)-C3 (dynamical friction)  shows that gravitational scattering slightly increases the number of subhaloes at $r\le 0.35 r_{\rm vir}$, whereas if comparing C3--C4 (no dynamical friction, no scattering) we observe that the effects of dynamical friction can be neglected at all radii.  \\

 In the lower panel of Fig.~\ref{fig:r} we have analysed the effects of the mass resolution on the spatial distribution of
substructures. We have
carried out calculations where the minimum initial mass of merging substructures was above two different thresholds $m_1=10^8
h^{-1}M_\odot$ (calculation C6) and $m_2=10^7 h^{-1}M_\odot$ (calculation C1); in other words, we vary the substructure mass resolution, denoted here as $m_1$,
$m_2$, where $m_2=m_1/10$.  For determining the present distribution we also choose those substructures with masses above $m_1$
(dotted line) and $m_2$ (full line) located within the virial radius.  The resulting cumulative profiles are compared against the
fitting expression obtained by Gao et al. (2004) from large cosmological N-body simulations, finding that the semi-analytic curves
presented here agree extremely well with the N-body results. Only for $r\le 0.2 r_{\rm vir}$ does the abundance obtained
semi-analytically become smaller than that predicted by the formula of those authors. However, this problem is likely due to their
selection of the fitting expression, since a similar mismatch also appears in their fit (see their Fig.~11).  As Gao et al. (2004)
claim, the mass resolution scarcely affects the distribution of substructures, only in the innermost region of the host halo is
the abundance of substructures slightly suppressed when reducing the mass resolution.

\subsubsection{Mass function}\label{sec:mass}
\begin{figure}
\vspace{10.5cm} \includegraphics{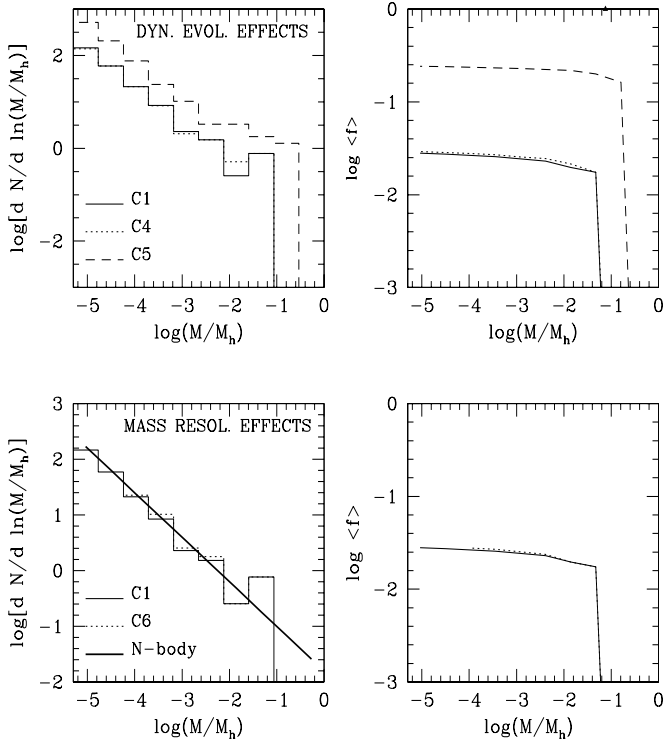}
\caption{Left column: Substructure mass function at $z=0$. Right column: Cumulative mass enclosed in substructures normalised to the host halo mass (for brevity, {\em mass fraction}). Upper-left panel shows the curves resulting from the implementation of dynamical friction and substructure scattering in eq.~(\ref{eqn:f_total}) (full line) and those where those forces were ignored (dotted line). For comparison, the dashed line accounts for the initial mass function (i.e. switching off mass loss). In the lower-left panel we plot the mass function for two different mass resolutions $m_1=10^8 h^{-1}M_\odot$ (dotted line) and $m_2=10^7 h^{-1}M_\odot$ (full line).}
\label{fig:m}
\end{figure}

We repeat the analysis of Section~\ref{sec:dist} for the substructure mass function. In Fig.~\ref{fig:m} we show in the upper-left
panel the curves resulting from calculations where the interaction forces (i.e. dynamical friction and substructure-substructure
scattering) were switched off (dotted line, calculation C4) and on (full line, calculation C1). As we can see, the differences are practically negligible on all
mass scales.  The dashed lines shows the resulting mass function if keeping constant the mass of substructures during their
evolution (calculation C5). In that case, substructures as massive as $M=0.25 M_{\rm h}(z=0)=5\times 10^{11} h^{-1}M_\odot$ can be found in the final
sample, whereas the total number of substructures is around a factor 3 larger than in calculations where the mass loss
scheme is implemented. Moreover, it is remarkable that both curves present a similar slope, which is due to the fact that, in
average, all substructures lose a similar mass fraction (see Section~\ref{sec:mevol}).

In the upper-right panel, we plot the averaged ratio between the mass enclosed in substructures and the host halo mass (which we
denote as $\langle f\rangle$) as a function of the substructure mass. This panel shows that the present population of
substructures represents only 2.8\% of the mass of a Milky Way-like halo. This result is in very good agreement with that of van den
Bosch, Tormen \& Giocoli (2004), who find $\langle f\rangle= 2.5\%$ for haloes with $M_{\rm h}(z=0)=2\times 10^{12}h^{-1}M_\odot$.  We
find that switching off mass loss leads to a mass fraction of $\langle f\rangle\simeq 25\%$, so that the mass lost by
substructures contributes in a significant amount to the host halo mass.  We note that the averaged mass fraction found here is
considerably smaller than that plotted in the lower-right panel of Fig.~\ref{fig:merger} (which results in $\langle f\rangle
\simeq 0.44$), since for this calculation we excluded all substructures at $r>r_{\rm vir}$.  Like van den Bosch, Tormen \& Giocoli
(2004), we find that most of the present day mass fraction originates from substructures with $M/M_{\rm h}>10^{-2}$ if implementing mass
loss (note that the curves are remarkably flat for lower mass values). However, the largest contribution is shifted to
$M/M_{\rm h}>10^{-1}$ if no mass loss occurs, so that the mass enclosed in substructures with $M/M_{\rm h}<10^{-1}$ represents a tiny fraction
of the entire mass despite their larger abundance.

In the lower panels, the above analysis is repeated for calculations with a mass resolution ten times smaller. Additionally, we
reproduce the formula suggested by 
van den Bosch et al. (2004) (strong full line) to fit the mass
functions obtained from cosmological N-body calculations.  As in the above panels, we obtain practically no difference in the
final mass function. 

Lastly, we must remark the good agreement between the N-body mass function and that obtained semi-analytically here\footnote{The over-prediction in the number of substructures with $M>0.1M_{\rm h}$ is a result of the {\em universality} of the fit (in the sense that they do not include the host halo mass dependence) carried out by Gao et al. (2004). Numerical and semi-analytical calculations of mass functions (e.g van den Bosch et al. 2004 and references therein) show that, for $M_{\rm h}(z=0)=2\times 10^{12} h^{-1}M_\odot$, a high-mass cut-off appears at $M_{\rm cut}\simeq 10^{11}h^{-1}M_\odot$, very similar to that found here ($M_{\rm cut}=1.2\times 10^{11}h^{-1}M_\odot$).}.

\subsection{Mass evolution of substructures}\label{sec:mevol}
In this Section we analyse the effects that dynamical friction and substructure scattering induce on the mass evolution of dark matter substructures. Additionally, we also determine the dependence of the results on the mass resolution.

\begin{figure}
\vspace{10.5cm} \includegraphics{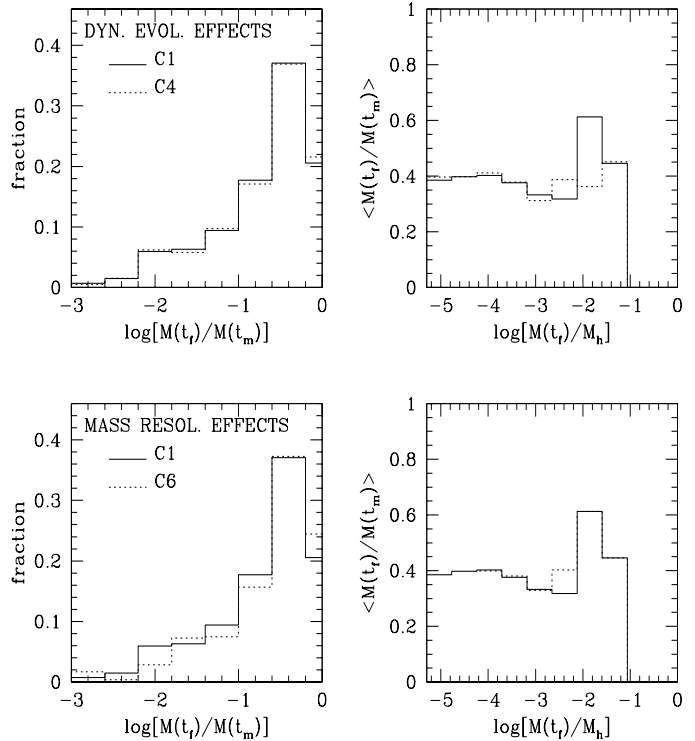}
\caption{Left column: Fraction of substructures as a function of the ratio of present mass to mass at time of accretion
($M(t_f)/M(t_{\rm m})$). Right column: Averaged mass loss as a function of the present mass $M(t_f)$. Upper raw: Effects of dynamical
evolution on mass loss. We compare the results from calculations with and without dynamical friction and gravitational scattering
(full and dotted lines, respectively). Lower raw: Effects of mass resolution. We compare the results from calculations with mass
resolution $m_2=10^7h^{-1}M_\odot$ (full lines) and $m_1=10^8h^{-1}M_\odot$ (dotted lines). For this plot, only substructures with
masses above the mass resolution (either $m_1$ or $m_2$, depending on the calculation) were included in the distribution.}
\label{fig:mloss}
\end{figure}

In Fig.~\ref{fig:mloss} we plot the fraction of substructures as a function of their mass decrease (left column) and the averaged mass loss as a function of the substructure present mass (right column). In the upper row we analyse the effects of dynamical evolution, whereas in the lower row we compare the results from calculations with different mass resolution.\\
In the upper-left panel we can see that dynamical friction and gravitational scattering induce slight changes on the mass loss process. The distribution peaks at $M(t_f)\sim 0.5$--$0.8 M(t_{\rm m})$. Strong mass loss events, where $m_2<M(t_f)<10^{-3}M(t_{\rm m})$ appear fairly rare. It is also interesting to note that a large number of substructures have lost a negligible fraction of their initial mass, either because they have been accreted relatively late or because their orbits keep them in the outer region of the halo, where tidal fields are weak.
In the lower-left panel we can also see that mass resolution scarcely affects the mass fraction lost by substructures during their evolution.\\
In order to infer a possible mass dependence in the mass loss process, we plot the averaged $M(t_f)/M(t_{\rm m})$ as a function of the present mass (right column). Interestingly, we find that substructures lose, on average, around 60\% of their initial mass for  $M(t_f)<10^{-2}M_{\rm h}(t_f)$, and 40\% for those with $M(t_f)>10^{-2}M_{\rm h}(t_f)$ ,although that difference might be caused by the relatively small number of massive substructures at $z=0$ (note that they represent less than 1\% of the total number of substructures in our sample, see Fig.~\ref{fig:m}). This result is mostly independent of dynamical evolution processes and mass resolution.

\section{Conclusions}\label{sec:concl}
In this paper, we have carried out semi-analytic calculations of the evolution of dark matter substructures in a Milky Way-like
halo. We have generated nine samples of substructures with spherical symmetry by following the technique of Cole et al. (2000),
which constructs the merger history of a halo back to high redshift.  Subsequently, our semi-analytical code calculates the orbit
of each substructure from the time of merging until the present epoch by solving its equation of motion.  Additionally, this code
implements a mass loss scheme that has been tested in previous papers and includes an analytical treatment of the tidal heating
induced by substructure-substructure collisions.

 We first use our code to carry out an illustrative example that quantifies the frequency and severity of two-body scatterings. We
inject into each of our samples two control substructures that initially follow a circular and an eccentric orbit. The encounter
parameters are recorded and statistically analysed. We find that overlapping encounters (those where the minimum distance is less
than $2 r_{\rm vir}$) are fairly rare, representing less than 2\% of the total number of impacts.  As a consequence, the
tidal heating experienced by substructures owing to substructure-substructure encounters induces a negligible expansion of their
mass profiles.

The main purpose of our semi-analytical investigation is to disentangle the effects of those processes acting on the evolution of substructures, which would be extremely expensive in terms of computational time if using large cosmological N-body calculations.  \\
After reproducing the most recent N-body results, we have separately analysed the forces that mainly drive the orbits of substructures, which are: (i) the force induced by the host halo potential, (ii) dynamical friction and (iii) substructure-substructure scattering.  Additionally, we have also varied the mass resolution of our calculations by selecting only those substructures with a mass at their merging time above two different minimum masses $10^8h^{-1}M_\odot$ and $10^7h^{-1}M_\odot$.
We find that:\\
(i) Neither dynamical friction nor substructure-substructure scattering induce strong variations on the present radial profile of substructures. Gravitational scattering slightly increases the number of substructures in the inner regions of the host halo, $r<0.35 r_{\rm vir}$, whereas the effects of dynamical friction can be neglected at all radii. The maximum increase in the cumulative number of substructures is less than 5\% if compared with the distribution where both terms were switched off.\\ 
(ii) The effects of those forces on the present mass function are negligible.

%{\bf JP: 
We have studied the orbital evolution of substructures by measuring their angular momentum each time-step. In contrast to the energy, the angular momentum is an adiabatic invariant that remains constant in varying potentials in the absence of external forces. We find that the most massive subhaloes suffer some degree of angular momentum loss, $\langle L(z=0)/L(t_{\rm m})\rangle \simeq 0.7$ for $M(z=0)\sim 0.1 M_{\rm h}(z=0)$. For less massive substructures, the averaged angular momentum remains practically unaltered since the time of their merging.
%}

%{\bf AJB: We have to be careful in making our conclusions here. While dynamical friction and scattering don't affect the subhalo spatial and mass distributions overall, I think they must make a difference to the spatial distribution of the most massive subhaloes. That could then be a problem as it will be these subhaloes which are selected when people perform observational studies of satellite galaxies.}

We have also determined how the mass evolution of substructures affects their present distribution. We find that mass loss effects are much more important than those induced by dynamical friction and scattering. Mass loss processes
decrease the maximum mass expected for substructures at the present day from $M_{\rm cut}=0.25 M_{\rm h}$ down to $M_{\rm cut}=0.06
M_{\rm h}$---values in fairly good agreement with those reported by Gao et al. (2004).  We find that the mass contribution of disrupted substructures
to the present day halo mass is around 22\%, 
%{\bf AJB: But this must depend on resolution?}
whereas only $0.025M_{\rm h}$ is still enclosed in form of bound substructures. 
The present mass fraction in subhaloes mostly originates from a relatively low number of massive substructures ($M\sim 0.1 M_{\rm h}[z=0]$). As a consequence, the mass fraction is barely dependent on the mass resolution. These results agree fairly well with those of van den Bosch et al. (2004). 
 Lastly we find that, on average, substructures lose around 40--60\% of their mass by the present day,
independent of their initial mass. This explains the low sensitivity of the mass function slope to mass loss processes.
 
We conclude, therefore, that the present distribution of substructures is mainly determined by the merger tree and the mass
evolution of these systems, whereas dynamical friction and gravitational scattering play a negligible role.

%This result has an interesting application for N-body cosmological simulations. It is well known that effects induced by dynamical
%friction and two-body interactions in galaxies are fairly sensitive to computational parameters. The small influence of those
%terms can probably explain the remarkable low sensitivity observed in N-body substructure distributions on, for example, the
%numerical code used to simulate galaxy formation and evolution, the number of particles and the spatial resolution, once the
%cosmological model is fixed (e.g Gao et al. 2004).

These results are also important for understanding the present mismatch between CDM predictions and observations of satellite
galaxies:\\ (i) {\bf The missing satellite problem}. The apparent overestimation in the number of satellite galaxies (e.g Klypin,
Kravtsov \& Valenzuela 1999) cannot be explained by evolutionary processes within the host galaxy. The suggestion that dynamical
friction or scattering may strongly enhance the disruption of substructures by driving them to the inner regions of the galaxy can
be discarded.\\ (i) {\bf The Holmberg effect}. The anisotropic distribution of satellite galaxies (Holmberg 1969, Zaritsky \&
Gonz\'alez 1999, Sales \& Lambas 2004), where most appear to follow nearly polar orbits out to 500 kpc, cannot be
explained by dynamical processes acting within the host halo.

Another interesting consequence of our results is that satellite
galaxies beyond $0.35r_{\rm vir}\simeq 100 h^{-1}$ kpc with masses
$M(z=0)<0.05 M_{\rm h} \simeq 10^{11}h^{-1}M_\odot$ have adiabatic
invariants that have not been altered by dynamical processes acting
within the host galaxy. Thus, these objects retain their orbital
properties which are determined {\em a priori} by the intergalactic
environment. Although we must keep in mind that observational surveys
are usually biased towards massive satellite galaxies, which may
present some degree of dynamical evolution (in particular we find that
subhaloes with $M(z=0)>0.05 M_{\rm h}$ show a decrease of their
initial angular momentum of around 25--30\% on average), observations
of satellite orbits should provide a useful tool to infer the
intergalactic dark matter distribution.

\section{Acknowledgements} 

AJB acknowledges support from a Royal Society University Research Fellowship.  JP thanks I. Trujillo for helpful discussions.

{}


\begin{thebibliography}{}
\bibitem[]{}Bacon D.J., Refregier A.R., Ellis R.S., 2000, MNRAS, 318, 625
\bibitem[]{}Benson A.J., Frenk C.S., Lacey C.G., Baugh C.M., Cole S., 2002a, MNRAS, 333, 156
\bibitem[]{}Benson A.J., Frenk C.S., Lacey C.G., Baugh C.M., Cole S., 2002b, MNRAS, 333, 177
\bibitem[]{}Benson A.J., Lacey C.G., Frenk C.S., Baugh C.M., Cole S., 2004, MNRAS, 351, 1215
\bibitem[]{}Binney J., Tremaine S., 1987, Galactic Dynamics. Princeton University Press, Princeton, New Jersey 
\bibitem[]{}Bullock J.S., Kravtsov A.V., Weinberg D.H., 2000, ApJ, 539, 517
\bibitem[]{}Colpi M., Mayer L., Governato F., 1999, ApJ, 525, 720 
\bibitem[]{}de Lucia G., Kauffmann G., Springel V., White S.D.M., Lanzoni B., Stoehr F., Tormen G., Yoshida N., 2004, MNRAS, 348, 333
\bibitem[]{}Gao L. S. D., White M., Jenkins A., Stoehr F., Springel V., 2004, astro-ph/0404589
\bibitem[]{}Ghigna S., Moore B., Governato F., Lake G., Quinn T., Stadel J., 2000, ApJ, 544, 616
\bibitem[]{}Gnedin O.Y., Hernquist L., Ostriker J.P., 1999, ApJ, 514, 109
\bibitem[]{}Gnedin O.Y., Ostriker J.P., 1999, ApJ, 513, 623
\bibitem[]{}Hernquist L., 1993, ApJS, 86, 389
\bibitem[]{}Holmberg E., 1969, Arkiv. Astr, 5, 305
\bibitem[]{}Kauffmann G., White S.D.M., Guiderdoni B., 1993, MNRAS, 264, 201
\bibitem[]{}King I.R., 1962, AJ, 67, 471
\bibitem[]{}Klypin A., Gottl\"ober S., Kravtsov A.V., Khokhlov A.M., 1999, ApJ, 516, 530
\bibitem[]{}Klypin A., Kravtsov A.V., Valenzuela O., 1999, ApJ, 522, 82
\bibitem[]{}Klypin A., Zhao H., Somerville R.S., 2002, ApJ, 573, 597
\bibitem[]{}Kneib J.-P., Hudelot P., Ellis R.S., Treu T., Smith G.P., Marshall P., Czoske O., Smail I., Natarajan P., 2003, ApJ, 598, 804
\bibitem[]{}Moore B., Ghigna S., Governato F., Lake G., Quinn T., Stadel J., Tozzi P., 1999, ApJ, 524, 19
\bibitem[]{}Navarro  J.F., Frenk C.S., White S.D.M., 1997, ApJ, 490, 493
\bibitem[]{}Pe\~{n}arrubia J., Kroupa P.,  Boily C.M., 2002, MNRAS, 333, 779
\bibitem[]{}Pe\~{n}arrubia J., 2003, PhD Thesis, {Universit{\"a}t }Heidelberg, Germany. {\tt http://www.ub.uni-heidelberg.de/archiv/3434}
\bibitem[]{}Pe\~{n}arrubia J., Just A., Kroupa P., 2004, MNRAS, 349, 747
\bibitem[]{}Percival W.J. et al., 2001, MNRAS, 327, 1297
\bibitem[]{}Piatek S., Pryor C., 1995, AJ, 109, 1071
\bibitem[]{}Prada P. et .al, 2003, ApJ, 598, 260
%\bibitem[]{}Prugniel Ph., Combes F., 1992, A\&A, 259, 25
\bibitem[]{}Sales L., Lambas D.G., 2004, MNRAS, 348, 1236
\bibitem[]{}Sand D.J. Treu T., Smith G.P., Ellis R.S., 2004, ApJ, 604, 88
\bibitem[]{}Somerville R.S., 2002, ApJ, 572, 23
\bibitem[]{}Spergel D.N. et al., 2003, ApJS, 148, 175
\bibitem[]{}Springel V., White S.D.M., Torman G., Kauffmann G., 2001, MNRAS, 328, 726
\bibitem[]{}Taylor J.E., Babul A., 2001, ApJ, 559, 716
\bibitem[]{}Taylor J.E., Babul A., 2004, MNRAS, 348, 811
\bibitem[]{} van den Bosch F., Lewis, G.F., Lake G., Stadel J., 1999, ApJ, 515, 50
\bibitem[]{}Vel\'azquez H., White S.D.M., 1999, MNRAS, 304, 254 
\bibitem[]{}Weldrake D.T.F., de Blok W.J.G., Walter F., 2003, MNRAS, 340, 12
%\bibitem[]{}Wahde M., Donner K.J., 1996, A\&A, 312, 431
\bibitem[]{}Zaritsky D., Gonz\'alez A.,1999, PASP, 111, 1508
\bibitem[]{}Zhao H., 2004, MNRAS, 351, 891
\end{thebibliography}
\end{document}